\documentclass{pasj00}
\draft

\begin{document}
\SetRunningHead{Author(s) in page-head}{Running Head}
\Received{****/**/**}
\Accepted{****/**/**}

\title{
A Chain of Dark Clouds in Projection Against the Galactic Center }

 \author{%
   Takahiro \textsc{Nagayama}\altaffilmark{1},
   Shuji \textsc{Sato}\altaffilmark{2}, 
   Shogo \textsc{Nishiyama}\altaffilmark{2,3}, 
   Yuka \textsc{Murai}\altaffilmark{1}, \\
   Tetsuya \textsc{Nagata}\altaffilmark{1}, 
   Hirofumi \textsc{Hatano}\altaffilmark{2},
   Mikio \textsc{Kurita}\altaffilmark{2}
   Motohide \textsc{Tamura}\altaffilmark{3},
   Yasushi \textsc{Nakajima}\altaffilmark{3}, \\
   Koji \textsc{Sugitani}\altaffilmark{4}, 
   Tomoharu \textsc{Oka}\altaffilmark{5}, 
   Yoshiaki \textsc{Sofue}\altaffilmark{6}, \\
   }
 \altaffiltext{1}{Department of Astronomy, Kyoto University, Sakyo-ku, Kyoto 606-8502, Japan}
 \altaffiltext{2}{Department of Astrophysics, Nagoya University, Furo-cho, Chikusa-ku, Nagoya 464-8602, Japan}
 \altaffiltext{3}{National Astronomical Observatory of Japan, Mitaka, Tokyo 181-8858, Japan}
 \altaffiltext{4}{Graduate School of Natural Science, Nagoya City University, Mizuho-ku, Nagoya 467-8501, Japan}
 \altaffiltext{5}{Research Center for the Early Universe and Department of Physics,\\ The University of Tokyo, 7-3-1 Hongo, Bunkyo-ku, Tokyo 113-0033, Japan}
 \altaffiltext{6}{Institute of Astronomy, The University of Tokyo, Mitaka, Tokyo 181-0015}
 \email{nagayama@kusastro.kyoto-u.ac.jp}

\KeyWords{xxxx:xxxx ......} 

\maketitle

\begin{abstract}
In the $J$, $H$, and $K_{S}$ bands survey of the the Galactic Center region over an area of \timeform{2D}  $\times$ \timeform{5D}, we have found many dark clouds, among which a distinguished chain of dark clouds can be identified with a quiescent CO cloud.
The distances of the clouds is estimated to be 3.2-4.2 kpc, corresponding to the Norma arm by our new method to determine distance to dark clouds using the cumulative number of stars against $J-K_{S}$ colors.
Adopting these estimated distances, the size is about $\sim$70 pc in length and the total mass of the cloud is 6$\times$10$^{4}$M$\solar$.
Three compact HII regions harbor in the cloud, indicating that star forming activities are going on at the cores of the quiescent CO cloud on the spiral arm.
\end{abstract}

\section{Introduction}
The central region of our Galaxy suffers large interstellar extinction, and images taken toward the Galactic Center (GC) in the optical wavelengths are generally featureless star fields with few stars.
Since the discovery of an infrared source corresponding to the GC by Becklin and Neugebauer (1968), the GC region has been observed extensively in the infrared, where interstellar extinction is one order of magnitude smaller than in the optical.
Succeeding to the larger-scale near-infrared (NIR) surveys of 2MASS (Skrutskie et al. 2006) and DENIS \citep{Epchtein94}, which have given a general view of the Galactic structure, we have carried out a $J$, $H$, and $K_{S}$ bands survey of the $\timeform{2D}$  $\times$ $\timeform{5D}$ region centered on the GC with 2 mag deeper sensitivity and 2 times finer resolution than those of 2MASS. 
In the course of the survey, we found numerous dark clouds silhouetted against the dense star fields.

Toward the GC, existence of many molecular clouds is well known from radio observations (Morris and Serabyn 1996 and references therein).
Their location in the line of sight is uncertain, but part of them, located on the near side of the GC, should appear as dark clouds in the near infrared images.  However, the correspondence of near infrared dark 
clouds to radio molecular clouds has not been studied well.

The distance to dark/molecular clouds is crucial for discussing their physical parameters associated with them.  
In radio observations, kinematic distances to molecular clouds are often derived from their radial velocity on the assumption that each cloud follows the circular Galactic rotation. 
However, since the Galactic rotation is perpendicular to our line of sight to the GC, we cannot determine the distance to the molecular clouds.
Moreover, non-circular components such as expanding rings make situations complicated.

Sofue (2006) proposed a method to determine the distance of CO clouds in the direction of GC from the gradient in the $l-v$ diagram, and determined the distance of CO clouds toward the GC.
This method is powerful for determining the distance to arm-like structures, which extend long in the $l$ direction, but unsuited for small CO clouds because the velocity gradient is not determined well.

In this paper, we report a chain of dark clouds found in our near infrared survey, which shows good positional and morphological agreements with a velocity component of $^{12}$CO ($J$=1-0) observed with the Nobeyama radio telescope (Oka et al. 1998).  
A new method to determine the distance to dark clouds from a diagram of $J-K_S$ color and cumulative star number, and the resultant distance to the chain of dark clouds are presented.

\section{Observation}
\subsection{NIR data}
Our near infrared survey toward the \timeform{2D}($b$) $\times$ \timeform{5D}($l$) region of the Galactic Center was carried out from 2002 to 2004.
The observation was made with SIRIUS, a simultaneous imager in the $J$ ($\lambda$=1.26\micron), $H$ ($\lambda$=1.63\micron) and $K_S$ ($\lambda$=2.14\micron) bands, covering an area 7$\arcmin.7 \times 7\arcmin.7$ with a pixel scale of 0\arcsec.45 on the IRSF 1.4-m telescope at Sutherland, South Africa.
The details of the instrument are described in \citet{Nagashima99} and \citet{Nagayama03}. 
We observed totally 891 of 7$\arcmin.7 \times 7\arcmin.7$ fields centered on the GC, with 10 times of 5 sec exposure.

The reduction of near-infrared array images was performed with IRAF (Imaging Reduction \& Analysis Facility).
Each image, after subtraction of the average dark frame, was divided by the normalized flat-field image. 
Then, the thermal emission pattern, the fringe pattern due to OH emission, and the slope pattern seen in the HAWAII arrays were subtracted from each frame with a median sky frame.
The photometry for point sources was performed with the DAOPHOT package \citep{Stetson87}, and calibrated with the standard star 9172 in Persson et al. (1998).
The 10$\sigma$ limiting magnitudes for the $J$, $H$, and $K_S$ bands are 17.1, 16.6, and 15.6 mag, respectively, and stars brighter than 10 mag in all the three band are saturated.
More details of our data reduction and photometries are described in \citet{Nishiyama05}.

\subsection{CO data}
$^{12}$CO ($J$=1-0, 115.271202 GHz) observations over the area \timeform{-1.5D} $\le l \le$ \timeform{+3.4D} and \timeform{-0.6D} $\le b \le$ \timeform{+0.6D} were carried out from 1995 February to April, and from 1996 March to May with the 45 m telescope of Nobeyama Radio Observatory by Oka et al. (1998).
The spatial resolution of the telescope was \timeform{15"} and the grid spacing was \timeform{30"}, resulting in an actual resolution on the map of \timeform{34"}. 
The details of the observation are described in \citet{Oka98}.
We reprocessed the data and acquired 88 velocity channel maps covering the velocity range of $V_{LSR}$ = $-220$ to $+220$ km s$^{-1}$, integrated over successive 5 km s$^{-1}$ widths.

\begin{figure}
  \begin{center}
    \FigureFile(80mm,120mm){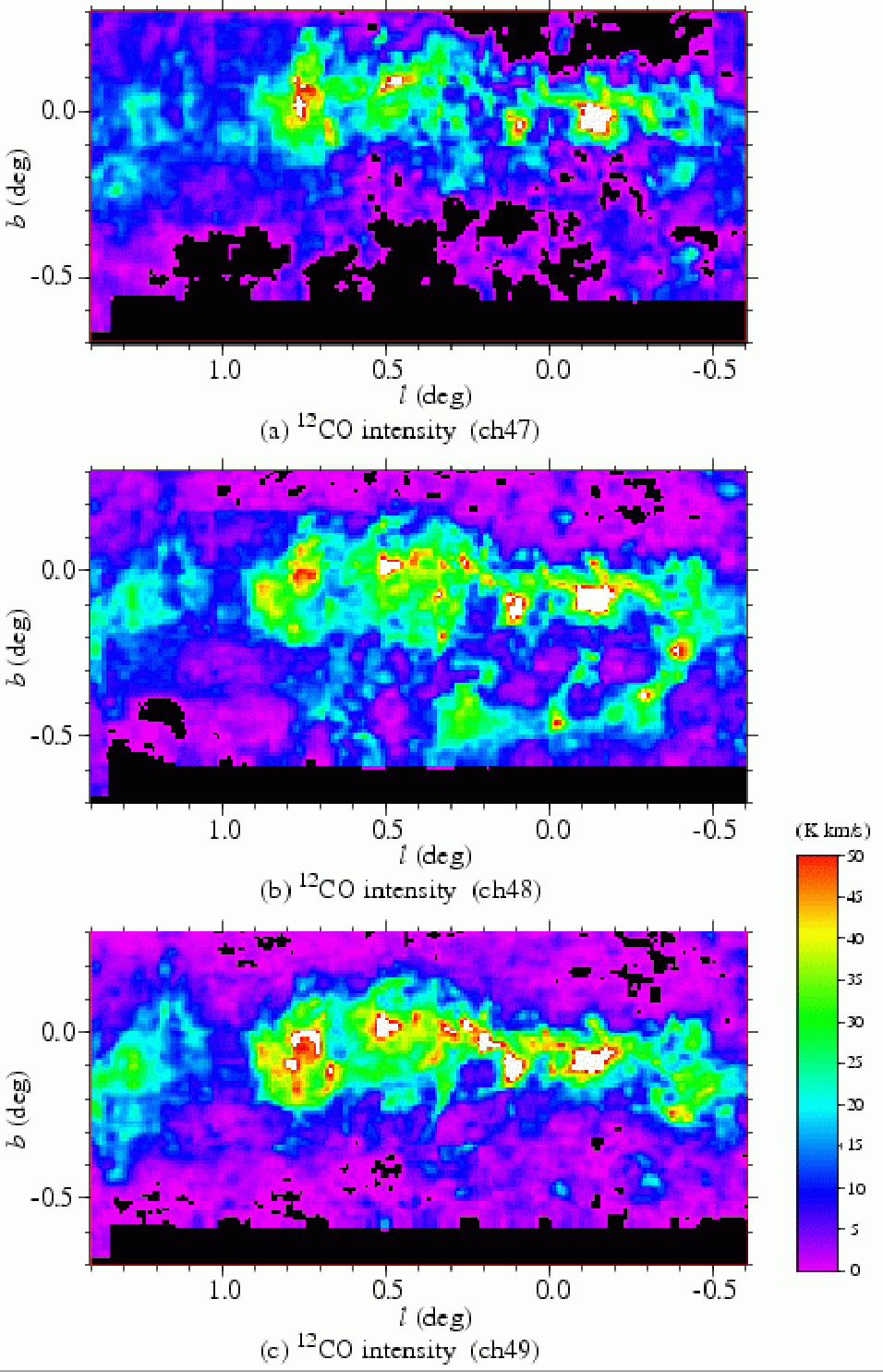}
  \end{center}
  \caption{$^{12}$CO intensity maps for (a)ch-47 (10$\le$ $V_{LSR}$ $\le$ 15 km s$^{-1}$), (b)ch-48 (15$\le$ $V_{LSR}$ $\le$ 20 km s$^{-1}$), (c)ch-49 (20$\le$ $V_{LSR}$ $\le$ 25 km$^{-1}$)}\label{fig:COmap}
\end{figure}

\section{Comparison of NIR images with CO maps}
We examined positional and morphological correspondences between the dark clouds seen in our survey and the $^{12}$CO emission.
By comparing our NIR images with the velocity-channel maps of the $^{12}$CO observation, we found that a filamentary high-intensity edge of  $^{12}$CO contours in the channel 48 (ch-48, $V_{LSR}=+15$ to $+20$ km s$^{-1}$) shows good correlation to a chain of dark clouds.
In Figs.1 (a)-(c), we show the $^{12}$CO intensity maps of ch-48 together with the adjacent channels (ch-47 and ch-49).
A filamentary structure extends from ($l$, $b$)$\sim$ (\timeform{-0.4D}, \timeform{-0.2D}) to (\timeform{+0.4D}, \timeform{-0.5D}) only in the map of ch-48.
Of particular interest is that the small velocity-width and velocity-gradient confined within single channel or a range less than 5 km s$^{-1}$.

For further comparison, we made a contour map of {I(ch-48)-(I(ch-47)+I(ch-49))}/2 superposed on the number density of stars brighter than 14.5 mag in the $Ks$ band (Fig. 2), in which the shape of $^{12}$CO contour conform with less star area in positional and morphological.
Therefore, the $^{12}$CO cloud only seen in ch-48 and the chain of dark clouds are presumably identical.

\begin{figure*}
  \begin{center}
    \FigureFile(180mm,120mm){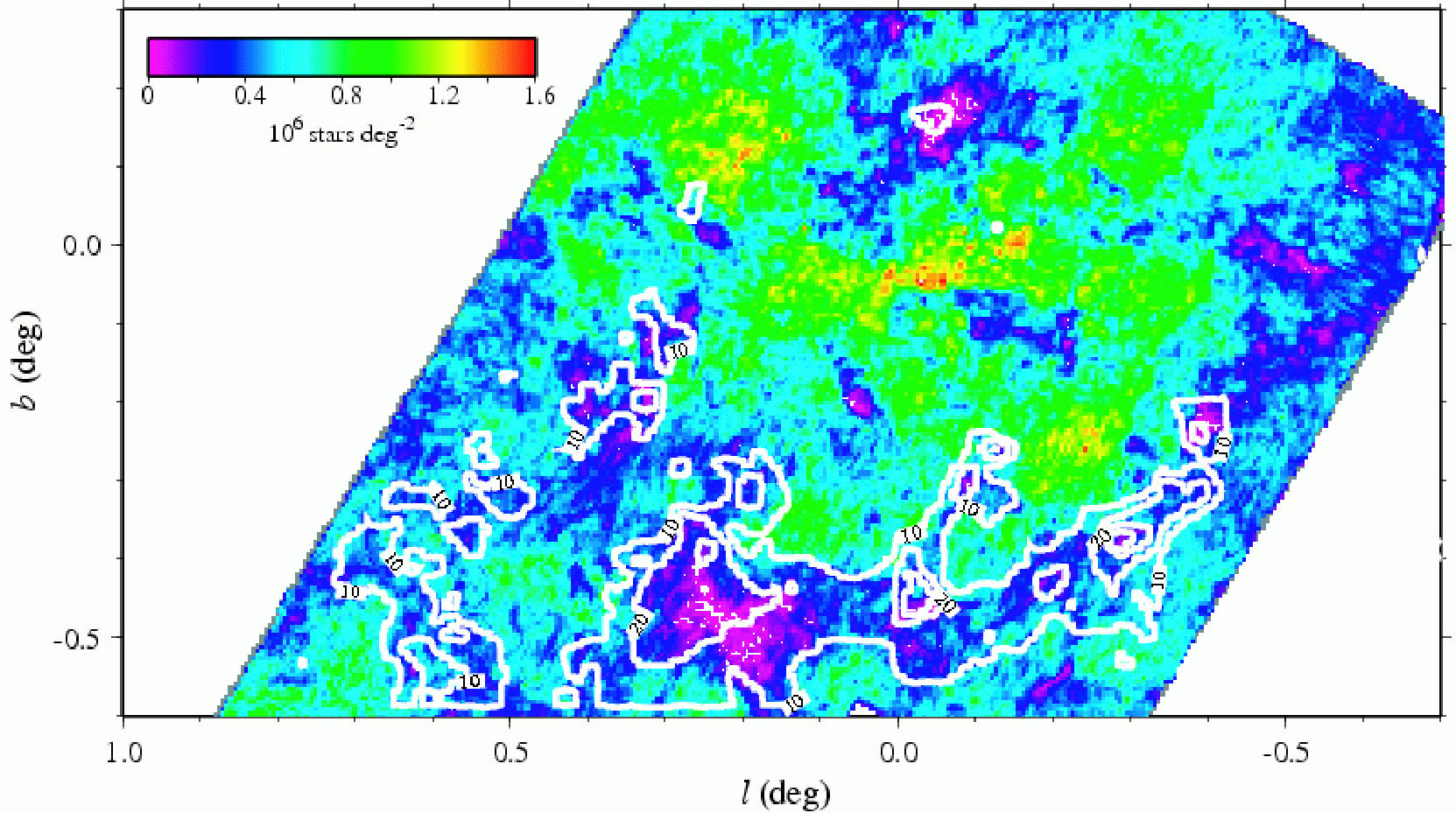}
  \end{center}
  \caption{$^{12}$CO-intensity excess (I(ch-48)$-$(I(ch-47)$+$I(ch-49))/2) contours superposed on the $Ks$ band star number density ($Ks$\textless14.5; color map). Contour levels are linear with a step of 10 K km s$^{-1}$. }\label{fig:COmap}
\end{figure*}

\section{Determination of the distance to the chain of dark clouds}
\subsection{A new method to determine distances to dark clouds}

\begin{figure}
  \begin{center}
    \FigureFile(80mm,120mm){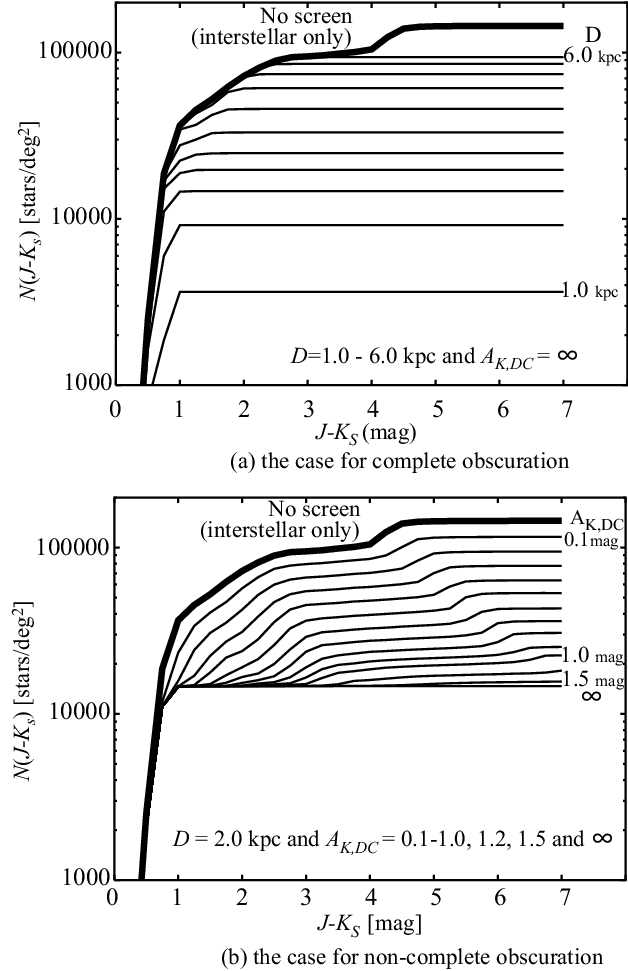}
  \end{center}
  \caption{The predicted $N(J-K_{S})$ versus $J-K_{S}$ color by changing screen parameters; (a) distance $D$, from 1.0 to 6.0 kpc (0.5 kpc step), and (b) extinction $A_{K,DC}$ from 0.1 to 1.0 mag (0.1 mag step), 1.2, 1.5 and $\infty$.  The thick (upper most) lines in each panel indicates the $N(J-K_{S})$ without a screen (interstellar extinction only). }

\end{figure}

Determination of the distance to dark clouds is a key in order to discuss the physical properties such as its real size and mass.  
We will demonstrate that the cumulative number of stars bluer than a certain $J-K$ color, $N(J-K)$, is a good indicator to estimate distances to dark clouds.
Provided that interstellar extinction and reddening increase monotonously with distance in a line of sight, the observed $J-K$ color of a star depends on the distance to the star.  
Therefore, the profile of $N(J-K)$ against various $J-K$ is determined by extinction in the line of sight, and conversely, we can estimate the extinction distribution along the line of sight, from the $N(J-K)$ profile.

The number of stars to be observed can be predicted using the infrared model of the Galaxy (Wainscoat et al. 1992).
We add an artificial screen with an extinction $A_{K_S, DC}$ at the distance $D$ as a dark cloud, besides the general interstellar extinction $A_{K_S}$ in the model.
Here, we assume that the thickness of dark clouds along the line of sight is thin enough compared with  the galactic scale; the thickness of screen is zero in our calculation.
We then calculate $N(J-K_{S})$ to be observed with our actual limiting magnitudes of $J=16.0$ and $K_{S}$=15.0 mag.
The detail of calculation is described in Appendix.

Fig. 3 shows predicted $N(J-K_S)$ toward $(l, b) = (-0.\timeform{3D}, -0.\timeform{3D})$ against $J-K_S$.
First, we discuss the case that the cloud is a complete screen and we cannot detect stars behind the cloud.
In Fig. 3 (a), the 10 lines ($D$ = 1.0 to 6.0 kpc with a step of 0.5 kpc) are shown, together with  the line without a screen.
The top thick line represents the case of no-screen with only interstellar extinction, while the 10 lines are for a complete screen with $A_{K_S, DC}=\infty$ in addition to interstellar extinction.
In these cases, $N(J-K_{S})$ is contributed by only foreground stars.

The $N(J-K_{S})$ profiles with the complete screen increase as same as the no-screen $N(J-K_{S})$ profile at small $J-K_{S}$ because of the foreground stars, but are detached from the no-screen line, beginning with the nearest $D$ case.
Because the detached points are different for different $D$, we can determine $D$ by comparing predicted and observed detached points.
After detached from the no-screen line, the increase of $N(J-K_{S})$ stops at $N_{fg}$, the number of foreground stars detectable with our sensitivity.
$N_{fg}$ is uniquely determined for $D$ in the cases of completely obscuring clouds.

Next, we discuss the case that the cloud does not obscure the stars behind completely.
In this case, certain number of stars behind the cloud are detected with extra extinction and reddening due to the cloud.
Fig. 3 (b) shows the case of $D=2$ kpc with $A_{K_S, DC}=0.1-1.0$ mag (0.1 mag step), 1.2, 1.5 and $\infty$ mag.
As we described in Fig 3 (a), $N(J-K_S)$ with $A_{K_S, DC}=\infty$ stays at $N_{fg}$  because background stars are undetectable.
In the case of $A_{K_S, DC}$=1.0 mag, all background stars are about 2 mag ($A_{K_S}=0.494E(J-K_S)$, Nishiyama et al. 2006a) reddened in $J-K_s$ by the screen.
Therefore, the increase of the $N(J-K_{S})$ with $A_{K_S, DC}$=1.0 stops at $N_{fg}$ at $J-K_S$=1, but $N(J-K_S)$ increases again from $J-K_S$=3 due to contribution of background stars.
Because the different $A_{K_S, DC}$ gives different width of $J-K_S$ stopping at $N_{fg}$, we can determine $A_{K_S}$ from it.

As we described using Fig. 3 (a) and (b), the $N(J-K_S)$ profiles reflects $D$ and $A_{K_S, DC}$, separately.
Therefore, we can determine the $D$ and $A_{K_S, DC}$ by comparing the observed $N(J-K_S)$ with the template parameterized by $D$ and $A_{K_S, DC}$.
Fig. 4 shows  simplified templates to determine $D$ and $A_{K_S, DC}$ for $(l, b) = (-0.\timeform{3D}, -0.\timeform{3D})$; a total of 13 lines for the cases of no dark screen (interstellar extinction only) and combination of $A_{K_S, DC}$ (0.5 and 1.0 mag) and $D$ (1, 2, 3, 4, 5 and 6 kpc) are shown.
Because the lines with various combination of $A_{K_S, DC}$ and $D$ are well separated from each other, we can assign an unique pair of $ A_{K_S, DC}$ and $D$ for the observed $N(J-K_S)$ profile of dark clouds.

\begin{figure*}
  \begin{center}
    \FigureFile(180mm,120mm){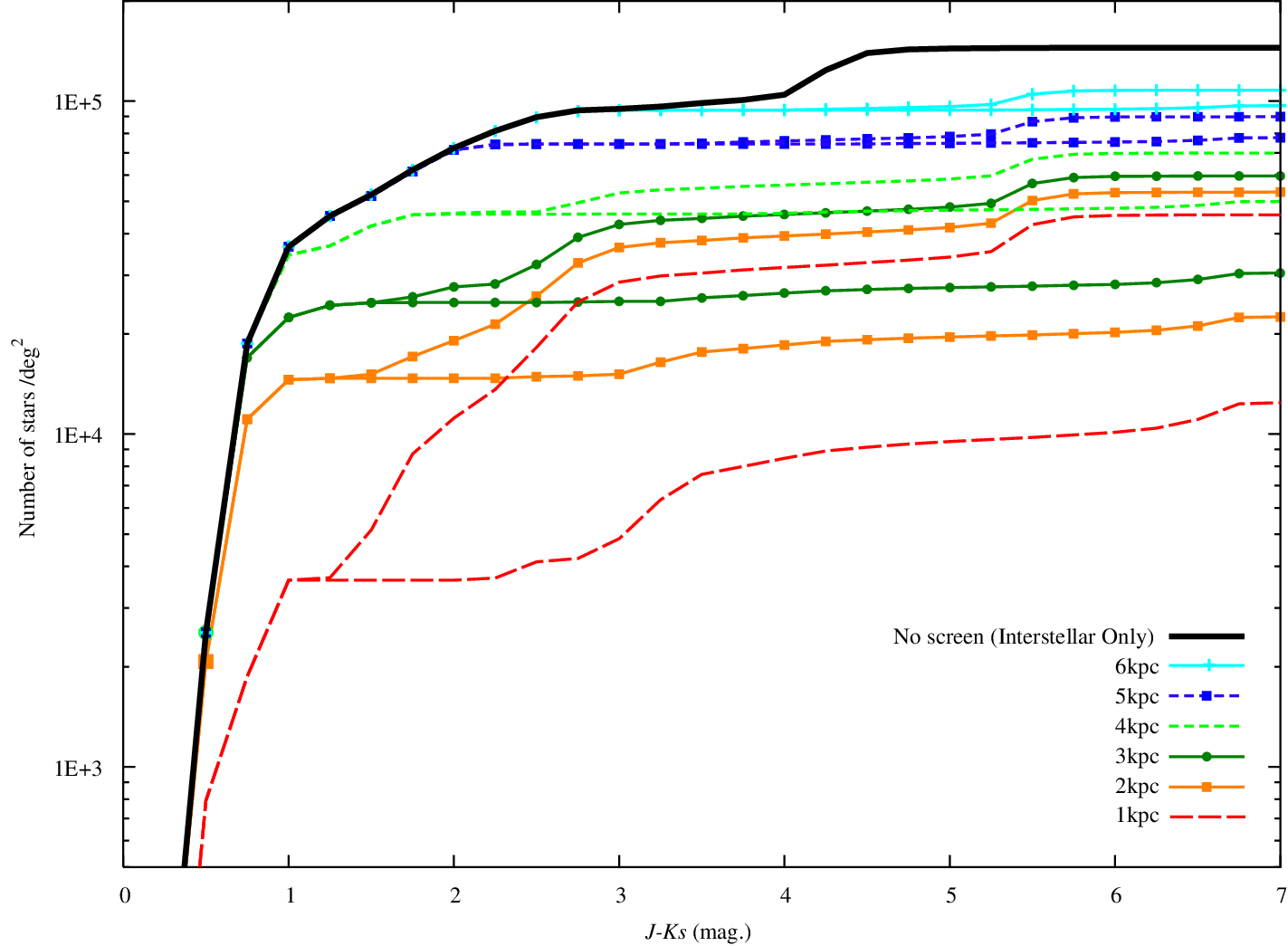}
  \end{center}
  \caption{A simplified temperate to determine $D$ and $A_{K_S,DC}$ using predicted $N(J-K_S)$ are shown. The lines with $A_{K_S}$ =0.5 and 1.0 mag located at $D$=1(red), 2(orange), 3(green), 4(light green), 5(blue), and 6(light blue) kpc are drawn. Upper and lower tracks of each color indicate $A_{K_S,DC}$=0.5 and 1.0 mag, respectively.}\label{fig:j-k-03-03}
\end{figure*}

\subsection{Comparison of our method with the Wolf diagram}

In the Wolf diagram, a traditional method to estimate the distance to dark clouds, cumulative number of stars observed brighter than magnitude $m$, $N(m)$, is adopted instead of $N(J-K)$.
The distance to the dark cloud is estimated from the apparent magnitude where the $N(m)$ profile for the dark cloud is detached from that for the comparison field.  
In reality, however, the detached point is very ambiguous due to the large dispersion in the absolute magnitudes of the stars as pointed out by Trumpler and Weaver (1953).  

By contrast, the intrinsic color of stars in the near infrared tends to be similar among almost all the spectral types; the $J-K$ of dwarfs and giants lie between 0.0 and 1.0 (Wainscoat et al., 1992).
The color excess $E(J-K)$ due to interstellar reddening is also expected up to  $\sim 4$ mag at the distance of the GC.  
Thus, the large change of $J-K$ compared with the ambiguity of intrinsic $J-K$ of stars enables us to determine the distance to dark clouds.

\subsection{Distance to  the chain of dark clouds}
Ten zones were selected around and in the chain of dark clouds : two reference fields (RF1 and RF2) and eight dark clouds (DC1 to DC8) (Fig. 5).
In order to confirm the reliability of the Galaxy model toward the GC region, we compared the observed and calculated $N(J-K_{S})$ toward the reference fields, RF1 and RF2.
The $N(J-K_{S})$ profiles for two reference fields are both in fairly good agreements (Figs. 6 (a) and (b)).
We therefore regarded the Galaxy model as reliable enough and adopt it for determination of the distances.

We calculated $N(J-K_S)$ for DC1 to DC8.
We tried to search for the best fit variables, $A_{K_S,DC}$ and $D$, with steps of 0.2 mag and 0.1 kpc, respectively.
The model lines with best fit $A_{K_S,DC}$ and $D$ for DC1-8 are shown together with the observed $N(J-K_S)$ histograms (Figs. 6 (c)-(j) ).
The best fit lines show good agreement compared to the RF1 and RF2, where the model lines show small discrepancies.
A possible reason that the fittings for RF1 and RF2 are worse than those for the dark cloud regions is the uncertainty of model for the inner-galaxy, which becomes insignificant when dark clouds are present.

The results for DC1-DC8, summarized in Table 1, show convergence to a specific distance of 3.2-4.2 kpc.    
Presumably, these dark clouds lie on a single chain of molecular clouds at a distance of $\sim$4 kpc with the extinctions range of $A_{K_S,DC}$ $\sim$ 0.4 to 1.0 mag.
The interstellar extinction in front of the clouds, $A_{K_S}$, is estimated to be 0.4 to 0.5 mag from the model calculation.
Most of the dark clouds previously studied are nearby, and the foreground extinction is negligibly small.
In contrast, the dark clouds studied in this paper are significantly farther, and therefore the cumulative interstellar extinction is comparable to $A_{K_S,DC}$. 

The CO cloud discussed in this paper was also identified with $V_{LSR}=$21 km s$^{-1}$ and $dv/dl=5.1$ km s$^{-1}$/deg (Sofue 2006).
Its distance was derived to be 4.7 kpc for $R_0=$8.0 kpc, which is now 4.4 kpc for $R_0=$7.5 kpc, and is slightly larger value than ours.

 The distance $3.2-4.2$ kpc is halfway to the GC at the distance of 7.5 kpc from us.
According to Vall\'ee (2008), a summary of previous studies of galactic structure, three arms, Sagittarius-Carina, Scutum-Crux and Norma arms intervene between the Galactic Center and us.
In Fig. 1 of Vall\'ee (2008), these three arms are placed to cross with the line of sight to the GC at approximately 0.8, 2.8 and 4.0 kpc from us, respectively. 
It is reasonable that the dark/molecular cloud is located around Norma arm, and rotates around the GC with a non-circular velocity components $V_{LSR}\sim+$20 km $s^{-1}$ toward the GC.

The Four-kpc molecular ring (Clemens, Sanders and Scoville 1988, Nakanishi and Sofue 2006) is also known as a component, crossing the line of sight to the GC. 
However, Vall\'ee (2008) claimed that the Four-kpc molecular ring is just assemblage of starting segments of four spiral arms, and we therefore leave it out of consideration.

\section{Discussion}
We estimated a projected length of the chain of dark clouds to be $\sim$70 pc and the typical thickness of the filaments is as thin as 0.1 degree, or about 7 pc, located at $\sim$30 pc off from the Galactic plane.
The gaseous filaments are themselves divided into finer structures, and several high-intensity knots are seen on the ridge of the $^{12}$CO contours.
By integrating the $^{12}$CO excess in the Fig. 2, the masses of each ridges are estimated to be an order of $\sim$10$^{4}$M\solar, and the total mass for the entire CO cloud results in 6$\times$10$^{4}$M\solar.
If we assume the same thickness in the line of sight (7 pc) as the apparent thickness and an extent of 70 pc, this total mass yields molecular gas density on the order of 18 M\solar pc$^{-3}$, or 4$\times$10$^{2}$ H$_{2}$ cm$^{-3}$.
This is a typical density for normal molecular clouds.

In the three high-intensity knots, we found four NIR sources, SF1-A, SF1-B, SF2, and SF3, with extents of $\sim10\arcsec$ to $\sim40\arcsec$ or 0.2-0.8 pc at 4 kpc.
Pseudo color images and position of them are shown in Fig. 7 and Table 2.
Mid infrared sources (MSX-PSC, \cite{Egan96}) are associated with all of them.
OH, H$_{2}$O, and/or CH$_{3}$OH masers (SIMBAD, or \cite{Caswell83a}, \cite{Caswell83b} and \cite{Menten91}) are also associated with SF1-A, SF1-B, and SF3 with $V_{LSR}$=$+15$ to $+23$ km s$^{-1}$, quite consistent to $V_{LSR}$ of $^{12}$CO.
Both their sizes and  $K_{S}$ magnitudes in 2MASS Point Source Catalog (9.15, 7.67, and 7.80 for SF1-A, SF2, and SF3) are consistent with those of typical compact HII regions at a distance $\sim$4 kpc.
We infer that the dark/molecular cloud discovered in this survey are molecular clouds in the Norma spiral arm at 4 kpc from us.

\begin{figure}
  \begin{center}
    \FigureFile(80mm,80mm){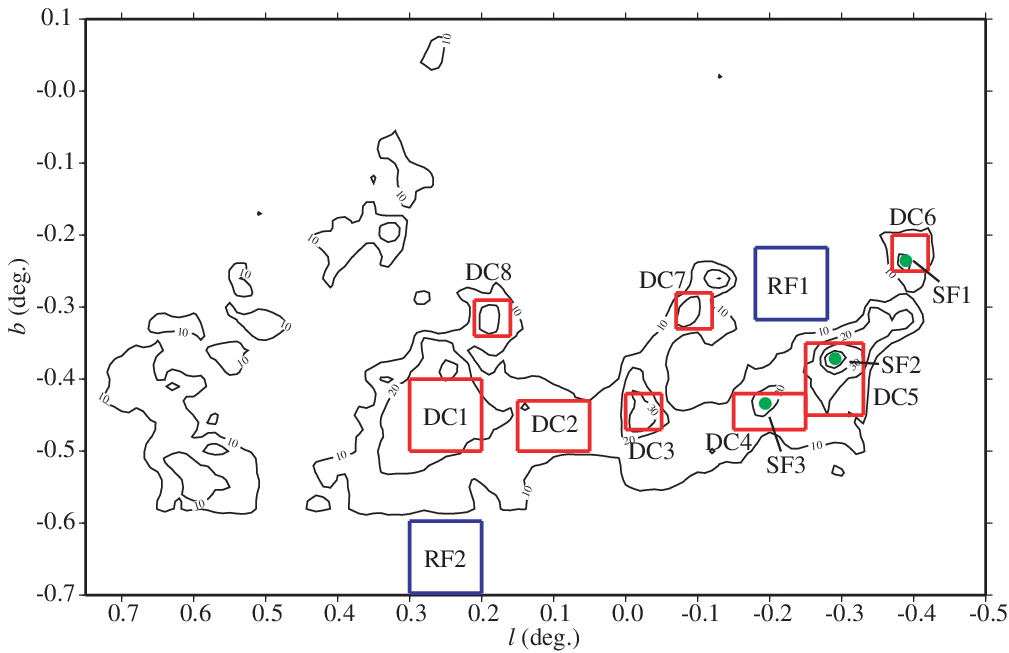}
  \end{center}
  \caption{Positions of RF1-2 and DC1-8 in the galactic coordinate, together with $^{12}$CO contour. }
  \label{fig:dc_rf_pos}
  \end{figure}

\begin{figure*}
  \begin{center}
    \FigureFile(160mm,120mm){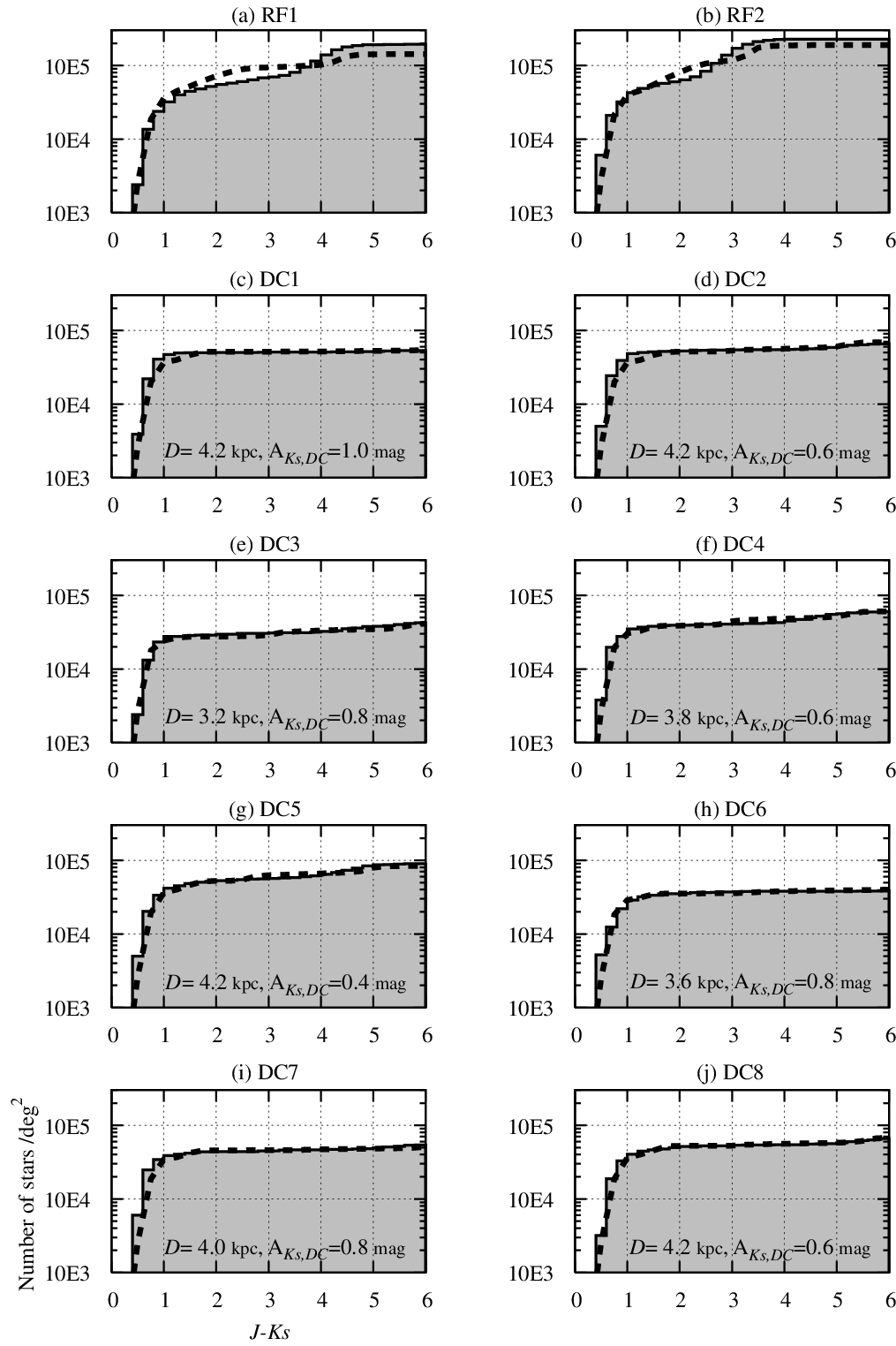}
  \end{center}
  \caption{Cumulative star-number histogram against $J-Ks$ toward two reference fields(RF1 and RF2) and eight DCs (DC1-8) are drawn. The dashed line indicates the predicted number from the Galaxy model. For RF1 and RF2, they are calculated with no  screen (interstellar extinction only). For DC1-8, the best-fit combination of the $D$ and $A_{K_S}$ are drawn. The best-fit parameters are written at bottom in each panel. }
\label{fig:model_comp_dc1-8_J-K}
\end{figure*}

\section{Summary}
We have found a number of dark clouds in our near infrared survey in the direction of the Galactic center region over an area of \timeform{2D} $\times$ \timeform{5D}.
From the agreement between the infrared and CO morphologies, we have identified a chain of the dark clouds located around the Norma spiral arm with a single CO molecular cloud lying at the distance of $\sim$4.0 kpc.
The chain of dark cloud harbors compact HII regions at the cores of the cloud in the CO spiral arm.

$ $

We wish to thank the staff of the South African Astronomical Observatory for their kind support during our observation and Dr. H. Okuda for his careful reading of the manuscript.

The IRSF/SIRIUS project was financially supported by the Sumitomo foundation and the Grant-in-Aid for Scientific Research on Priority Area (A) No. 10147207 and No. 10147214 of Ministry of Education, Culture, Sports, Science, and Technology (MEXT) of Japan.

This work is supported by the Grant-in-Aid for the 21st Century COE "Center for Diversity and Universality in Physics" from MEXT of Japan.

IRAF is distributed by the National Optical Astronomy Observatories, which are operated by the Association of Universities for Research in Astronomy, Inc., under cooperative agreement with the National Science Foundation.

This research has made use of the SIMBAD database, operated at CDS, Strasbourg, France

\appendix
\section{ The Galaxy model: Stellar populations with interstellar extinction}

 The numbers of stars detectable in the $J$, $H$, and $Ks$ bands have been estimated on the basis of the Galaxy model of Wainscoat et al. (1992).
It comprises geometrically and physically realistic representation of the Galactic disk, bulge, halo, spiral arms, and a molecular ring.
They represent each of the distinct Galactic components by space densities of 87 spectral-type stars with scale heights and absolute magnitudes at $B, V, J, H, K$, 12 and 25 \micron.

In our calculation, three components, an exponential disk, a bulge, and a molecular ring, were taken into account.
We adopted 7.5 kpc as $R_0$, the distance between the GC and us (Eisenhauer et al. 2005; Nishiyama et al. 2006b) instead of 8.5 kpc used in Wainscoat et al.(1992), which leads to a reduction factor 0.88 for the associated parameters.
We also took only 29 spectral types of main sequences and giants from Table 2 of Wainscoat et al.(1992) into calculation.

The stellar flux is subject to general interstellar extinction.
The amount of general interstellar extinction between us and distance $D_{obj}$, $A_{D_{obj}}$, is calculated as
\begin{equation}
    A_{D_{obj}} = \int_0^{D_{obj}} \gamma_{GC} \exp(-\frac{r}{h_{r}}) \exp(-\frac{|z|}{h_{z}}) dD,
\end{equation}
Here, $r$ and $z$ are the distance from the Galactic Center on the Galactic Plane and the distance from the Galactic Plane, and $h_{r}$ and $h_{z}$ are scale length of $r$ and $z$, adopting 3080 pc and 100 pc, respectively. 
$\gamma_{GC}$ is an amount of extinction per unit length at the Galactic Center, and it was obtained as follows.

First, we obtained the averaged extinction of red clump stars in the RF1 and RF2 ($A_{K_S}$=1.7 and 1.2, respectively), and assumed them as the total amount of general interstellar extinction to the Galactic Center ($D_{obj}$=7.5kpc).
The detail of method to obtain extinction is described in Nishiyama et al.(2005).
Next, we substituted $A_{D_{obj}}=1.7$ for RF1 and 1.2 for RF2 into the above equation, and derived $\gamma_{GC}$=0.76 for RF1 and 0.71 for RF2.
We adopted 0.73 mag/kpc, the average of the above two values, as $\gamma_{GC}$ in this work.

We calculated the apparent magnitude of 29 spectral types of stars every 10 pc bin along the given line of sight, and integrated the star number detectable by the given limiting magnitude and color ($J\le16$ and $K_{S}\le15$, 85\% detection rate of artificial stars in the most crowded region of our observation)  over the distance of 20 kpc.
We consider their sum to be the total number of stars to be detected.

\begin{table}
  \caption{Positions of DC1-8 and RF1-2}\label{tab:dc_rf_pos}
  \begin{center}
    \begin{tabular}{lllcc}
       \hline
       ID & $l$ (deg) & $b$ (deg) & $D$ (kpc) & $A_{Ks,DC}$ (mag) \\
       \hline
       DC1 & $+$0.250 & $-$0.450 &  4.2  & 1.0  \\
       DC2 & $+$0.100 & $-$0.465 &  4.2  & 0.6  \\
       DC3 & $-$0.025 & $-$0.445 &  3.2  & 0.8  \\
       DC4 (SF3) & $-$0.200 & $-$0.445 &  3.8  & 0.6  \\
       DC5 (SF2) & $-$0.290 & $-$0.400 &  4.2  & 0.4  \\
       DC6 (SF1) & $-$0.395 & $-$0.225 &  3.6  & 0.8  \\
       DC7 & $-$0.095 & $-$0.305 &  4.0  & 0.8  \\
       DC8 & $+$0.185 & $-$0.315 &  4.2  & 0.6  \\
       \hline
       RF1 & $-$0.230 & $-$0.270 & & \\
       RF2 & $+$0.250 & $-$0.650 & & \\
       \hline
    \end{tabular}

  \end{center}
\end{table}

\begin{table}
  \caption{Positions of SF1-A, SF1-B, SF2, SF3}\label{tab:sf_pos}
  \begin{center}
    \begin{tabular}{ccc}
       \hline
       ID & $l$ (deg) & $b$ (deg) \\
       \hline
       SF1-A & 359.615 & $-$0.244 \\
       SF1-B & 359.654 & $-$0.241 \\
       SF2 & 359.716 & $-$0.374 \\
       SF3 & 359.370 & $-$0.456 \\
       \hline
    \end{tabular}
  \end{center}
\end{table}

\begin{figure}
  \begin{center}
    \FigureFile(80mm,80mm){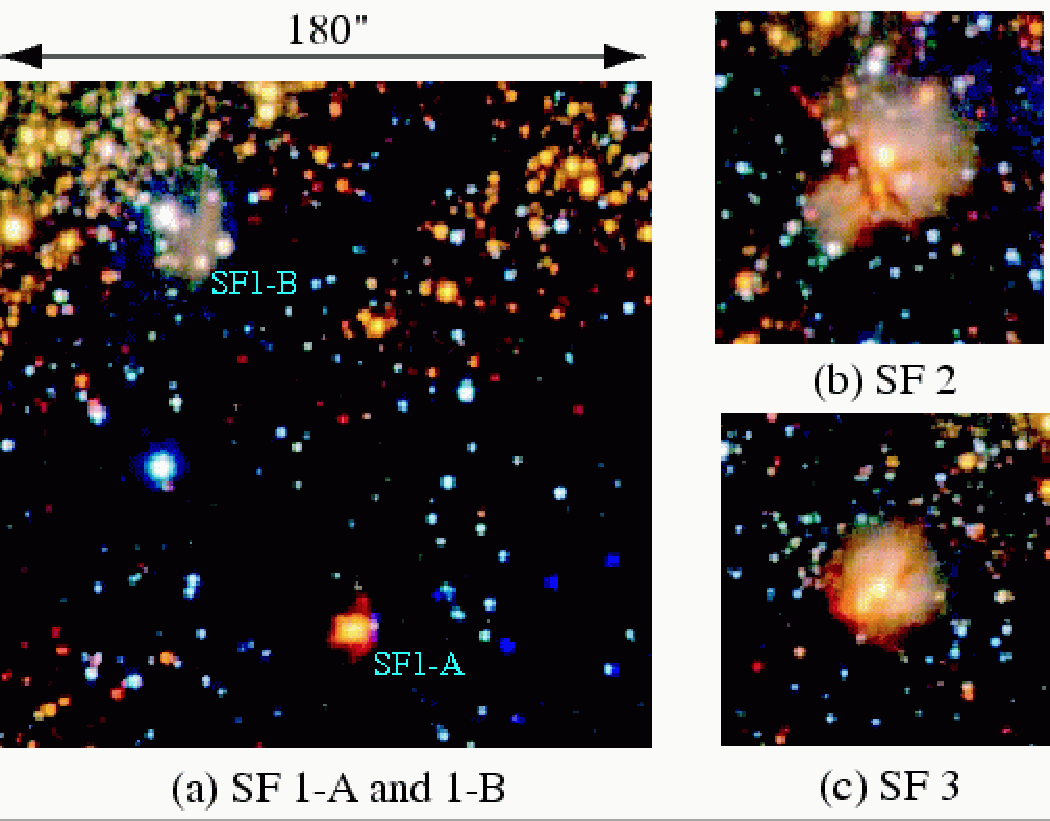}
  \end{center}
  \caption{Composite color ($J$: blue, $H$: green, and $Ks$: red) images of SF1, 2, and 3. North is up, east is left. Scale is the same among them.}\label{fig:SF1-3}
\end{figure}

\end{document}